\documentclass[twocolumn,showpacs,preprintnumbers,amsmath,amssymb,floatfix]{revtex4}

\usepackage{graphicx}
\usepackage{dcolumn}
\usepackage{bm}
\usepackage{subfigure} 

\begin{document}

\preprint{}

\title{Scenarios of heterogeneous nucleation and growth studied by cell dynamics simulation}

\author{Masao Iwamatsu}
\email{iwamatsu@ph.ns.musashi-tech.ac.jp}
\affiliation{
Department of Physics, General Education Center,
Musashi Institute of Technology,
Setagaya-ku, Tokyo 158-8557, Japan
}%


\date{\today}

\begin{abstract}
The dynamics of phase transformation due to homogeneous nucleation has long been analyzed using the classic Kolmogorov-Johnson-Mehl-Avrami (KJMA) theory.  However, the dynamics of phase transformation due to heterogeneous nucleation has not been studied systematically even though it is vitally important technologically.  In this report, we study the dynamics of heterogeneous nucleation theoretically and systematically using the phenomenological time-dependent Ginzburg-Landau (TDGL)-type model combined with the cell dynamics method.  In this study we focus on the dynamics of phase transformation when the material is sandwiched by two supporting substrates. This model is supposed to simulates phase change storage media.  Since both homogeneous and heterogeneous nucleation can occur simultaneously, we predict a few scenarios of phase transformation including: homogeneous-nucleation regime, heterogeneous-nucleation regime, and the homogeneous-heterogeneous coexistence regime.  These predictions are directly confirmed by numerical simulation using the TDGL model.  The outcome of the study was that the KJMA formula has limited use when heterogeneous nucleation exist, but it could still give some information about the microscopic mechanism of phase transformation at various stages during phase transformation.

\end{abstract}

\maketitle

\section{Introduction}
\label{sec:level1}

The dynamics of phase transformation by nucleation and subsequent growth of a nucleus of a stable phase is a very old problem, which has been studied for more than half a century~\cite{Christian,Kolmogorov,Johnson,Avrami,Price} from a fundamental point of view as well as from technological interests.  However, it plays a vital role today, for example, in information technology (IT) of phase-change optical data storage media.  For example, rewritable (RW) compact disks and digital video disks (DVD) consist of Te based alloys.  The data unit "bit" is switched between 0 and 1 by laser heating between the amorphous and the crystalline state~\cite{Weidenhof,Ruitenberg,Kalb2}.  The crystallization occurs through the heterogeneous nucleation at the surface of the disk.  Therefore, the data transfer rate of such a phase change storage media depends critically on the time scale of amorphous to crystalline phase transformation by the heterogeneous nucleation~\cite{Weidenhof,Kalb2}.  

The dynamics of phase transformation is usually described by the Kolmogorov-Johnson-Mehl-Avrami (KJMA) kinetics rule~\cite{Christian,Kolmogorov,Johnson,Avrami,Price}. According to this rule the fraction of transformed volume follows a straight line in the so-called Avrami-plot and the dynamics is characterized by the slope called the Avrami exponent.  The time scale of phase transformation is determined from the {\it incubation time} that is the waiting time before the start of the phase transition, and the {\it growth time} that is the time necessary to accomplish the phase transformation. The incubation time depends on the transient nucleation rate~\cite{Shneidman1}, while the growth time depends both on the steady-state nucleation rate and the growth rate of a nucleus.

Theoretically, the validity of the KJMA kinetics is well understood~\cite{Christian,Kolmogorov,Johnson,Avrami} and tested~\cite{Shneidman3,Iwamatsu4} for homogeneous nucleation, but it is not always obvious for heterogeneous nucleation ~\cite{Christian}.  Various modeling and analytical calculations seem to have limited use~\cite{Weinberg}.  Therefore, theoretical modeling and a computer simulation method that can handle both homogeneous and heterogeneous nucleation on the same footing are highly desirable.  

In the previous paper~\cite{Iwamatsu4,Iwamatsu1}, we adopted the cell dynamics method~\cite{Oono} to study KJMA kinetics of phase transformation within the framework of time-dependent Ginzburg-Landau~\cite{Valls} or phase field~\cite{Castro} model. The cell dynamics method can successfully describe the classical picture of homogeneous nucleation as well as the interface-limited growth~\cite{Chan} of nuclei.  Therefore, the whole process of KJMA kinetics ~\cite{Christian} of homogeneous nucleation and growth can be simulated on the same footing using cell dynamics.  Then, the time scale of transformation that consists of growth time and incubation time can be studied. The same story should be effective for heterogeneous nucleation and growth as well.  Similar studies of heterogeneous nucleation using the phase field model have appeared very recently~\cite{Emmerich,GranasyA,GranasyB}, but they paid more attention to the early stage of grwoth by spreading. 

In this paper, we will use the cell dynamics method with thermal noise to study heterogeneous nucleation and growth under various conditions and settings in the TDGL model in a unified manner. In Section II we present a short review of the classical picture of nucleation and growth.  We also discuss a few expected scenarios of heterogeneous nucleation present on the surface.  In Section III, we will present the cell dynamics method with thermal noise for studying heterogeneous nucleation and growth.  The surface field induces heterogeneous nucleation.  Section IV is devoted to the results of numerical simulations.  There, we pay attention to the Avrami exponent, the incubation time, and the nucleation rate which play dominant roles in the transformation time scale. Finally Section V is devoted to the conclusion.

\section{Dynamics of heterogeneous nucleation and growth}

The time evolution of the volume fraction $f_{\rm homo}$ of transformed volume for homogeneous nucleation, for example in two-dimensional infinite system, predicted by KJMA (Kolmogorov-Johnson-Mehl-Avrami)~\cite{Christian,Kolmogorov,Johnson,Avrami,Shneidman1} theory is as follows:
\begin{equation}
f = 1 - \exp\left(-\frac{\pi}{3}I_{s}v_{s}^{2}\left(t-t_{\rm inc}\right)^{3}\right).
\label{eq:3-0}
\end{equation}
which can be generalized as
\begin{equation}
f_{\rm homo} = 1 - \exp\left(-\left(\frac{t-t_{\rm inc}}{\tau_{\rm gr}}\right)^{m}\right),
\label{eq:3-1}
\end{equation}
where $m$ is the so-called Avrami exponent, $t_{\rm inc}$ the incubation time, and $\tau_{\rm gr}$ is the growth time, which is determined from the steady state nucleation rate $I_{s}$ and the steady state interfacial velocity $v_{s}$ through
\begin{equation}
\frac{1}{\tau_{\rm gr}}=\left(\frac{\pi}{m}I_{s}v_{s}^{2}\right)^{1/m}.
\label{eq:3-2}
\end{equation}
Then, the double logarithm of (\ref{eq:3-1}) gives a straight line
\begin{equation}
\log_{10} \left(-\ln (1-f_{\rm homo})\right)=m\log_{10} \left(t-t_{\rm inc} \right) + \mbox{const}.
\label{eq:3-3}
\end{equation}
between $\log \left(-\ln (1-f_{\rm homo})\right)$ and $\log \left(t-t_{\rm inc} \right)$ which is the so-called Avrami plot.  The slope gives the Avrami exponent $m$.

According to the theory of Shneidman and Weinberg~\cite{Shneidman1}, if we take into account the time-dependent transient nucleation rate for $I_{s}$ as well as the size-dependent interfacial velocity for $v_{s}$ and the finite size of the critical radius $R_{*}$ of nucleus, the incubation time $t_{\rm inc}$ is given by:
\begin{equation}
t_{\rm inc}\simeq \tau\frac{W_{*}}{k_{\rm B}T}-\tau.
\label{eq:3-4}
\end{equation}
where $\tau\simeq R_{*}/v_{s}$ is the time lag due to the finite size $R_{*}$ of the
critical nucleus, $W_{*}$ is the energy barrier to form the critical nucleus and $T$ is the absolute temperature

In classical nucleation theory~\cite{Christian,Kelton,Oxtoby,Iwamatsu2}, the steady-state homogeneous nucleation rate $I_{s}$, that is the number of critical nuclei which appear per unit time and unit volume, is usually given by the activation form
\begin{equation}
I_{s}\propto \exp\left(-\frac{W_{*}}{k_{\rm B}T}\right),
\label{eq:3-5}
\end{equation}
where $W_{*}$ is the nucleation barrier of critical nucleus that appeared in (\ref{eq:3-4}). 

In order to derive KJMA formula (\ref{eq:3-1}) we have used the following assumption~\cite{Shneidman1}:
\begin{enumerate}
\renewcommand{\labelenumi}{(\alph{enumi})}
\item the system is infinite and no boundary effect exists.
\item the nucleation is homogeneous.  Therefore, the nuclei form at random sites.
\item the steady state interfacial velocity $v_{s}$ is constant or weakly (power-law) time-dependence.
\item the steady state nucleation rate $I_{s}$ is constant or weakly (power-law) time-dependent.
\end{enumerate}
In an ideal case when the steady state velocity $v_{s}$ and the steady state nucleation rate are strictly constant, the Avrami exponent $m$ in (\ref{eq:3-1}) is given by 
\begin{equation}
m = d + 1
\label{eq:3-9a}
\end{equation}
for $d$-dimensional system.  Then $m=3$ for $d=2$ dimensional system.  Incidentally, when the fixed number of nuclei exist initially and continuous nucleation is suppressed (no nuclei are added), which is called site-saturation case~\cite{Christian}, we have
\begin{equation}
m=d
\label{eq:3-9b}
\end{equation}
Therefore, the Avrami exponent becomes smaller in the site-saturation case.


The first and the second conditions (a) and (b) are violated, however, for heterogeneous nucleation.  Furthermore, the energy barrier $W_{*}$ is lowered by the amount~\cite{Oxtoby,Shneidman4}
\begin{equation}
W_{*}\rightarrow W_{*}f(\theta)
\label{eq:3-10}
\end{equation}
with
\begin{equation}
f(\theta)=\frac{\theta-\sin\theta\cos\theta}{\pi}\leq 1
\label{eq:3-11}
\end{equation}
for the circular meniscus, where $\theta$ is the contact angle~\cite{Oxtoby}. For the complete or partial wetting condition $\theta=0$, the wetting layer spread over 
the surface which means that heterogeneous nucleation occurs without
crossing the energy barrier since $W_{*}f(\theta)=0$ from (\ref{eq:3-11}).  Furthermore
the incubation time becomes negative from (\ref{eq:3-7}).  Even in the partial wetting~\cite{Cahn,deGennes} condition $0<\theta<\pi/2$, the energy barrier is lowered $W_{*}f(\theta)<W_{*}$ and heterogeneous nucleation can occur more easily than homogeneous nucleation. The incubation time for heterogeneous nucleation also becomes shorter than for the homogeneous nucleation~\cite{Shneidman4}. 

When heterogeneous nucleation can occur near the substrate, we may expect coexistence of heterogeneous nucleation near the substrate and of homogeneous nucleation in the bulk. Then the time evolution of the volume fraction $f$ of transformed volume would be given approximately by the superposition of the two contributions:
\begin{equation}
f = f_{\rm homo} + f_{\rm hetero}
\label{eq:3-13}
\end{equation}
where $f_{\rm hetero}$ is the transformed volume by homogeneous nucleation (\ref{eq:3-1}) and $f_{\rm hetero}$ is the one by heterogeneous nucleation.  The transformed
volume fraction increases due to the simultaneous occurrence of both homogeneous and heterogeneous nucleation.

The latter contribution can be estimated by considering when the homogeneous nucleation
is completely suppressed.  Then, we may expect that nucleation occurs only at the substrate as heterogeneous nucleation and new phase growth from substrate as a traveling wave with front velocity $v_{s}$.  Then, the transformed volume is simply proportional
to the time:
\begin{equation}
f_{\rm hetero} \propto v_{s}t
\label{eq:3-14}
\end{equation}
since the front velocity $v_{s}$ depends only on the under cooling $\epsilon$ of the
material from (\ref{eq:3-9}) and does not depend on the existence of the substrate.  In this case, the usual KJMA formula (\ref{eq:3-1}) cannot be used.

Summarizing, we may expect a few phase transformation scenarios:
\begin{enumerate}
\renewcommand{\labelenumi}{(\alph{enumi})}
\item The homogeneous nucleation regime where heterogeneous nucleation can be neglected.  The phase transformation is governed by homogeneous nucleation in the bulk.  This situation occurs when the substrate is no-wet ($\theta=\pi$).
\item The heterogeneous nucleation regime where homogeneous nucleation can be neglected.  This can occur when the temperature $\xi_{0}^{2}$ is low or the undercooling $\epsilon$ is low and the substrate is wet ($\theta<\pi$) because the homogeneous nucleation rate will be low and would not be visible.
\item The coexistence regime where both homogeneous and heterogeneous nucleation can exist.  This can occur when the temperature $\xi_{0}^{2}$ is high or the undercooling $\epsilon$ is large and the substrate is wet ($\theta<\pi$), then the homogeneous and the heterogeneous nucleation rate as well as the interfacial velocity could be high.
\end{enumerate}
In those cases, the popular KJMA kinetic rule given by (\ref{eq:3-1}) is not applicable.  We have to analyze the transformed volume fraction $f$ directly.

\section{Cell Dynamics Simulation for Heterogeneous Nucleation and Growth}
\label{sec:sec2}

We use the standard time-dependent Ginzburg Landau (TDGL)~\cite{Valls} model to study the dynamics of heterogeneous nucleation.  The cell dynamics simulation method is used to solve the TDGL equation numerically.  This method digitizes the time and space of standard TDGL~\cite{Valls} evolution equation
\begin{equation}
\frac{\partial \psi}{\partial t}=-\frac{\delta \mathcal F}{\delta \psi},
\label{eq:2-1}
\end{equation}
where $\delta$ denotes the functional differentiation, $\psi$ is the {\it non-conserved} order parameter, and $\mathcal F$ is the free energy functional. This free energy is written as the square-gradient form
\begin{equation}
{\mathcal  F}[\psi]=\frac{1}{2}\int \left[D(\nabla \psi)^{2}+h(\psi)\right]{\rm d}{\bf r}. 
\label{eq:2-2}
\end{equation}
The local part $h(\psi)$ of the free energy functional $\mathcal  F$ determines the bulk phase diagram and the value of the order parameter in equilibrium phases.  

This TDGL equation (\ref{eq:2-1}) is loosely transformed into a space-time discrete cell dynamics equation~\cite{Oono}:
\begin{equation}
\psi(t+1,n)=F[\psi(t,n)]+\xi(t,n),
\label{eq:2-3}
\end{equation}
where the time $t$ is discrete and an integer, and the space is also discrete and is expressed by the integral site index $n$.  

The thermal noise $\xi(t)$ is added to (\ref{eq:2-1}) in (\ref{eq:2-3}) which is related to the absolute temperature $T$ from the fluctuation-dissipation theorem as
\begin{equation}
\left<\xi(t,n)\xi(t',n')\right>=k_{\rm B}T\delta_{n,n'}\delta_{t,t'}.
\label{eq:2-4}
\end{equation}
In this paper, we will use a uniform random number ranging from $-\xi_{0}$ to $+\xi_{0}$~\cite{Oono}.  Then, the parameter $\xi_{0}^{2}$ is proportional to the absolute temperature:
\begin{equation}
\xi_{0}^{2} \propto T,
\label{eq:2-5}
\end{equation}
or the temperature $T$ is included through the thermal noise $\xi_{0}$.

Since the cell dynamics method is invented not to simulate the mathematical TDGL partial differential equation (\ref{eq:2-1}) accurately but to simulate and describe the global dynamics directly~\cite{Oono}, it certainly cannot simulate the high-frequency fluctuation in space and time.  Therefore the thermal fluctuation eq.~(\ref{eq:2-5}) cannot take into account the short-time fluctuation.  However, such a fluctuation will be unimportant since the thermal fluctuation plays role mainly in the birth of nucleus and will play secondary role in time-evolution of nucleus. 

The mapping $F$ is given by
\begin{equation}
F[\psi(t,n)]=-f(\psi(t,n))+D\left[\ll\psi(t,n)\gg-\psi(t,n)\right],
\label{eq:2-6}
\end{equation}
where
\begin{equation}
f(\psi)=dh(\psi)/d\psi
\label{eq:2-6x}
\end{equation}
is the derivative of the local part $h(\psi)$ in (\ref{eq:2-2}), and the definition of $\ll\cdots\gg$ for the two-dimensional square grid is given by~\cite{Oono}
\begin{equation}
\ll\psi(t,n)\gg=\frac{1}{6}\sum_{i=\mbox{nn}}\psi(t,i)
+\frac{1}{12}\sum_{i=\mbox{nnn}}\psi(t,i),
\label{eq:2-7}
\end{equation}
where ``nn'' denotes nearest neighbors and ``nnn'' next-nearest neighbors.  In this report, we consider only the two-dimensional square grid to avoid heavy numerical work. 

We use the map function $f(\psi)$ directly obtained from the free energy $h(\psi)$~\cite{Iwamatsu4,Iwamatsu1} from (\ref{eq:2-6x}) instead of the standard $\tanh$ form originally used by Oono and Puri~\cite{Oono}, which is essential for studying the subtle nature of nucleation and growth when one phase is metastable and another is stable.

In order to study heterogeneous nucleation, we consider the situation
where the upper and lower side of material is sandwiched by substrates.
This situation simulate heterogeneous nucleation observed in Te based alloy of
rewritable (RW) compact disks and digital video disks (DVD)~\cite{Weidenhof,Ruitenberg,Kalb2}.  The effect of the
substrate, which induces heterogeneous nucleation, is represented by 
the surface field, which has been used to study the 
wetting~\cite{Cahn,deGennes} and heterogeneous nucleation~\cite{Talanquer}.
Here, we consider a $x-z$ two-dimensional system, and assume that the material is sandwiched by two lines parallel to the $x$-axis at $z=0$ and $z=d$.  

\begin{figure}[htbp]
\begin{center}
\includegraphics[width=0.8\linewidth]{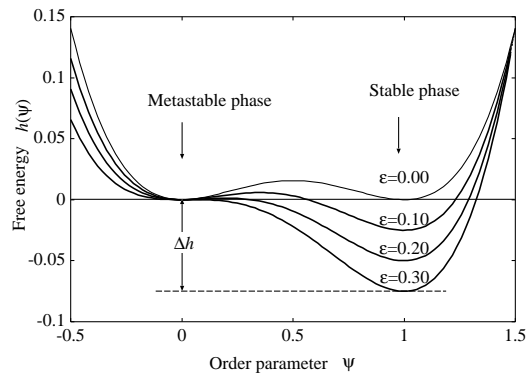}
\end{center}
\caption{
Model double-well free energy $h_{0}(\psi)$ defined by (\ref{eq:2-9}) that can realize two-phase coexistence when $\epsilon=0$.  The undercooling parameter $\epsilon$ determines the free energy difference $\Delta h$ between the depth of two wells.  The phase with $\psi_{m}=0$ is metastable while the one with $\psi_{s}=1$ is stable.  The spinodal occurs when $\epsilon=1/3=0.33\dots$
}
\label{fig:1}
\end{figure}

The local part of the free energy $h(\psi)$ we use~\cite{Iwamatsu4,Iwamatsu1}
consists of two parts:
\begin{equation}
h(\psi)=h_{0}(\psi)+h_{s}(\psi)
\label{eq:2-8}
\end{equation}
where the bulk free energy $h_{0}(\psi)$ is given by
\begin{equation}
h_{0}(\psi) = \frac{1}{4}\psi^{2}(1-\psi)^{2} + \frac{3}{2}\epsilon\left(\frac{\psi^{3}}{3}-\frac{\psi^{2}}{2}\right).
\label{eq:2-9}
\end{equation}
This free energy is shown in Fig. \ref{fig:1}, where one phase at $\psi_{m}=0$ is metastable while another phase at $\psi_{s}=1$ is stable. The free energy difference $\Delta h$ between the stable phase and the metastable phase is determined from the parameter $\epsilon$:
\begin{equation}
\Delta h = h(\psi_{m}=0)-h(\psi_{s}=1)=\frac{\epsilon}{4}.
\label{eq:2-10}
\end{equation}
Therefore, $\epsilon$ represents the undercooling which could be measured from melting point. We will use the terminology {\it undercooling} to represent $\epsilon$.
The metastable phase at $\psi_{m}=0$ becomes unstable when $\epsilon=1/3=0.33\dots$, which defines the spinodal.  

\begin{figure*}[htbp]
\begin{center}
\includegraphics[width=0.8\linewidth]{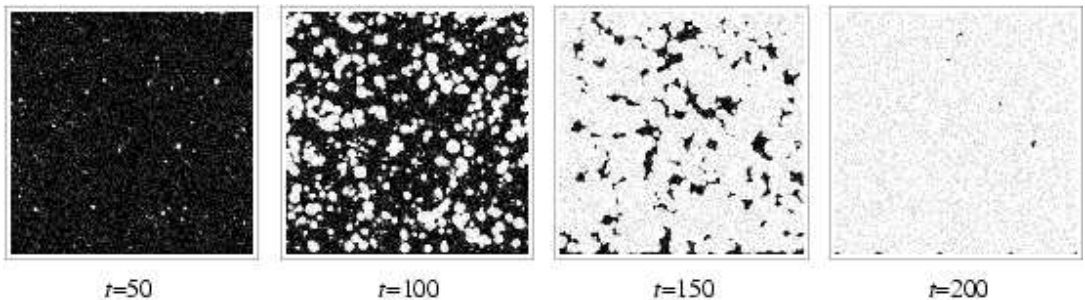}
\end{center}
\caption{
A typical evolution pattern of the phase transformation for 512$\times$512 system sandwiched by a neutral substrates with $\gamma_{u}=0$ at the top and a non-wet substrate with $\gamma_{l}=-0.4$ at the bottom calculated from our cell dynamics code for a medium undercooling $\epsilon=0.2$ and at a lower temperature $1/\xi_{0}^{2}=60$.  Nucleation occurs predominantly within the bulk and the nuclei grow isotropically preserving an almost circular shape.  At a later stage ($t=200$) a small amount of metastable materials are left as bubbles of metastable phase (black) attached to the lower substrate.}
\label{fig:2}
\end{figure*}

The surface free energy $h_{s}(\psi)$ is given by~\cite{Cahn,Talanquer}
\begin{equation}
h_{s}(\psi) = -\left(\gamma_{l}\delta(z) + \gamma_{u}\delta(z-d)\right) \psi
\label{eq:2-11}
\end{equation}
where $\gamma_{l}$ denotes the short-ranged surface field of the lower substrate at $z=0$ and $\gamma_{u}$ denotes one of the upper substrate at $z=d$. $\delta(x)$ is the usual
delta function.  At the surface of the substrate free energy is lowered by 
the amount $\gamma_{l}\psi_{l}$ at the lower substrate and $\gamma_{u}\psi_{u}$ at the upper substrate, where $\psi_{l}$ and $\psi_{u}$ are the order parameters in contact with the upper and the lower substrates.  Therefore, the substrate prefers to be wetted by the new stable phase with $\psi_{s}=1$ when $\gamma_{l,u}>0$, and the nucleation of the stable phase occurs preferentially at the surface of the substrate.  Then, heterogeneous nucleation could be promoted at the substrate.  

Due to the presence of the boundary at $z=0$ and $z=d$, the TDGL equation (\ref{eq:2-1})
should be augmented with the boundary condition:
\begin{eqnarray}
2D\left(\frac{\partial \psi}{\partial z}\right)_{l} &=& -\gamma_{l}
\nonumber \\
2D\left(\frac{\partial \psi}{\partial z}\right)_{u} &=& \gamma_{u}
\label{eq:2-12}
\end{eqnarray}
where subscript $l$ means that the derivative should be calculated at the surface 
of the lower substrates at $z=0$ and $u$ means the upper substrate surface at $z=d$. 

A similar boundary condition with $\gamma=0$ was used by Castro~\cite{Castro} and later by Gr\'an\'asy~\cite{GranasyB}, which corresponds to the free boundary condition.  It is also possible to fix the order parameter, for example, by setting $\psi_{l}=1$ at the boundary. This strategy was also adapted by several researchers to study the heterogeneous nucleation~\cite{GranasyB,Athreya} and wetting~\cite{Luo}.  We use the boundary condition (\ref{eq:2-12}) because it is directly related to the contact angle of the popular theory of wetting~\cite{deGennes,Cahn}.  It should be noted, however, this contact angle does not mean naive geometrical angle, rather the angle represents wettability derived from the thermodynamic consideration.  Therefore, the boundary condition used by S\'emoroz~\cite{Semoroz} is merely a phenomenological condition and does not represent any thermodynamic meaning.


In cell dynamics code, (\ref{eq:2-12}) is transformed into a discrete form.  If we use symbolically the index $n=0$ as the bottom of surface at $z=0$ and $n=N$ as the top of the surface at $z=N$, the boundary conditions for the phase-field $\psi(t,x,z)$ are given by
\begin{eqnarray}
\psi(t,x,0)&=& \psi(t,x,1)+\frac{\gamma_{l}}{2D}, \nonumber \\
\psi(t,x,N)&=& \psi(t,x,N-1)+\frac{\gamma_{u}}{2D},
\label{eq:2-13}
\end{eqnarray}
which takes into account the effect of the short-range field $\gamma_{u,l}$ of 
the surface.  In this case, however, the function $f$ in the map function $F$ 
in (\ref{eq:2-6}) is given $f(\psi)=dh_{0}(\psi)/d\psi$ of the bulk because the effect
of the surface field is short-ranged. 
A similar condition was used by Marko~\cite{Marko} in his cell dynamics model to study the surface effect on the dynamics of the phase transformation of a conserved system.

By changing the magnitude of the interfacial energy $\gamma_{u}$ and $\gamma_{l}$, we can change the heterogeneous nucleation rate on the substrate.  It should be noted, however, that the growth velocity of the nucleus, even if they are heterogeneously nucleated at the substrate, depends only on the undercooling $\epsilon$ of the bulk~\cite{Iwamatsu4}.  The substrate acts only as the seed of the nuclei, therefore it only affects the nucleation rate at the substrate and does not influence the growth of the nuclei.


\section{Numerical Results and Discussion}


We used the cell dynamics code developed previously~\cite{Iwamatsu4}, and simulated phase transformation when the effects of substrate and of the heterogeneous nucleation exist. 
In our TDGL model (\ref{eq:2-2}) with (\ref{eq:2-9}), the energy barrier $W_{*}$ of the homogeneous nucleation of two-dimensional circular nucleus is given by
\begin{equation}
W_{*}=\frac{\pi\sigma^{2}}{\Delta h}\propto \frac{1}{\epsilon}
\label{eq:3-6}
\end{equation}
with its critical radius $R_{*}$ given by
\begin{equation}
R_{*}=\frac{\sigma}{\Delta h} \propto \frac{1}{\epsilon}
\label{eq:3-7}
\end{equation}
where $\Delta h$ is the free energy difference (\ref{eq:2-10}) between the metastable and the stable phase (Fig.~\ref{fig:1}).  The interfacial energy $\sigma$ between the stable and the metastable phase is given by
\begin{equation}
\sigma = \frac{1}{12}\sqrt{\frac{D}{2}},
\label{eq:3-8}
\end{equation}
which can be calculated from the interfacial profile calculated from the Ginzburg-Landau equation $\delta \mathcal F/\delta \psi=0$ for the order parameter $\psi$ at the two-phase coexistence ($\Delta h=0$)~\cite{Iwamatsu2}.  Then the $\epsilon$-dependence of the surface tension $\sigma$ could be ignored, and the energy barrier $W_{*}$ in (\ref{eq:3-6}) is mainly determined from the magnitude of undercooling $\epsilon$.  The energy barrier $W_{*}$ is high when the undercooling $\epsilon$ is low. 

\begin{figure}[htbp]
\begin{center}
\includegraphics[width=0.85\linewidth]{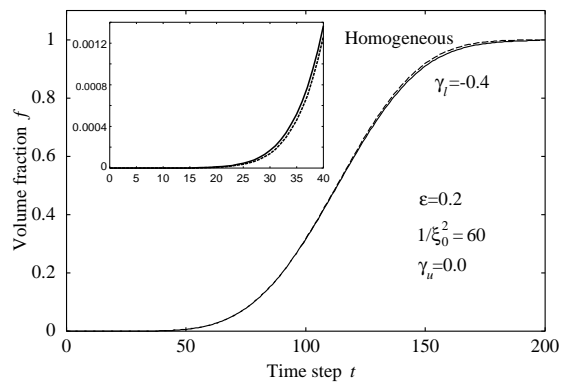}
\end{center}
\caption{
The time evolution of the volume fraction $f$ average over 50 samples for $512\times 512$ system when the material sandwiched by a neutral substrates with $\gamma_{u}=0$ at the top and a non-wet substrate with $\gamma_{l}=-0.4$ at the bottom (broken line) compared with the values when the system is infinite (homogenous nucleation). Standard deviations are not shown because they are too small to be visible in this scale. }
\label{fig:3}
\end{figure}

\begin{figure}[htbp]
\begin{center}
\includegraphics[width=0.85\linewidth]{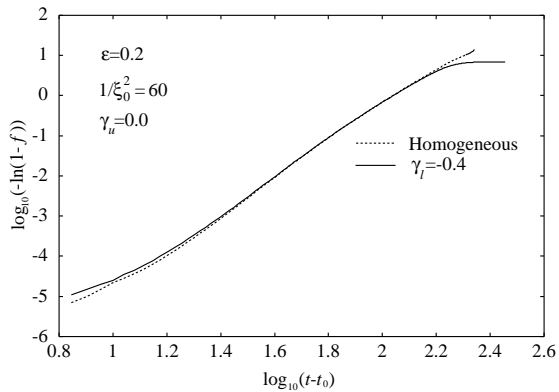}
\end{center}
\caption{
The Avrami plot corresponding to Fig. \ref{fig:3}. Again two curves are almost all in straight lines and the slope (Avrami exponent $m$) are almost the same. The Avrami exponents $m$ deduced from the linear portion of this Avrami plot are $m$=4.46 for $\gamma$=-0.4.  This value is very close to $m=4.45$ for homogeneous nucleation. }
\label{fig:4}
\end{figure}

\begin{figure*}[htbp]
\begin{center}
\includegraphics[width=0.8\linewidth]{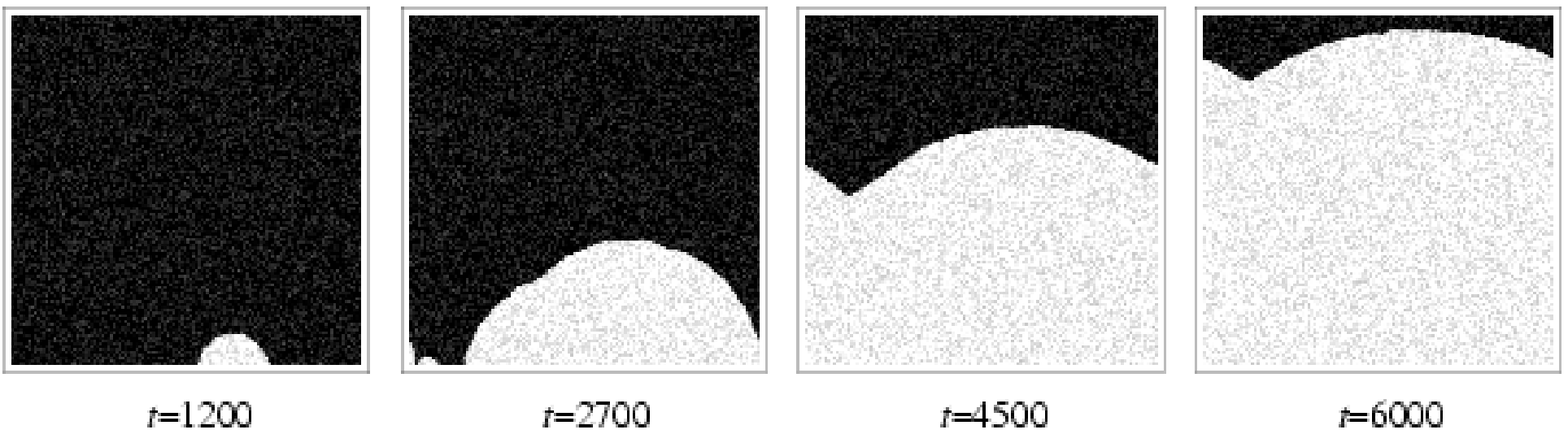}
\end{center}
\caption{
The same as Fig.~\ref{fig:2} but with a weakly wet substrate with $\gamma_{l}=+0.07$ at the bottom for the low undercooling $\epsilon=0.1$ and at a medium temperature $1/\xi_{0}^{2}=40$.  Homogeneous nucleation is suppressed and the nucleation occurs preferentially from the lower wet substrate. Only a very small number of nuclei formed at the substrate because of weak surface field, which leads to the very strong sample dependence of the evolution of the volume fraction $f$.}
\label{fig:5}
\end{figure*}

In our TDGL model, it has been shown that the nucleus grows by the interface-limited growth. The interfacial front velocity $v_{s}$ is almost constant and is roughly given by~\cite{Chan,Iwamatsu1,IwamatsuE}
\begin{equation}
v_{s}=\frac{1}{2}\sqrt{\frac{D}{2}}3\epsilon
\label{eq:3-9}
\end{equation}
using the undercooling $\epsilon$.  The deeper the undercooling, the higher the front velocity becomes. The interfacial velocity $v_{s}$ does not depend on the temperature.  Therefore, the temperature dependence of growth time $\tau_{\rm gr}$ in (\ref{eq:3-2}) comes solely from the temperature dependence of nucleation rate $I_{s}$ given by (\ref{eq:3-5}). Furthermore, the growth velocity of nuclei, which nucleate in the bulk and substrate, should be the same.

The contact angle $\theta$ is connected to the surface energy $\gamma$ through the Yound-Laplace equation~\cite{deGennes,Oxtoby}
\begin{equation}
\cos\theta = \frac{\gamma_{l,u}}{\sigma}\left(\psi_{s}-\psi_{m}\right)=\frac{\gamma_{l,u}}{\sigma}
\label{eq:3-12}
\end{equation}
as the surface free energies are given by $\gamma_{l,u}\psi_{s}$ and $\gamma_{l,u}\psi_{m}$ respectively at the lower and the upper substrates when the substrates are covered by the stable phase $\psi_{s}=1$ and the metastable phase $\psi_{m}=0$. 

Thermal noise is introduced as a uniform random number $r$ picked up from $-\xi_{0}\leq r \leq \xi_{0}$. The system size is fixed to $512\times 512$ since a larger system with $1024\times 1024$ was shown to give almost the same results~\cite{Iwamatsu4}.  Throughout this study, we set the value $D=1/2$.  Then the surface tension $\sigma$ between stable and metastable phase is $\sigma=1/24\simeq 0.042$.

\subsection{Homogeneous nucleation regime}

When the substrate is non-wet $\gamma<0$, heterogeneous nucleation is completely suppressed.  Then, phase transformation occurs through homogeneous nucleation and growth.  

Figure~\ref{fig:2} shows a typical pattern of evolution of a stable new phase (white) in the meta-stable old phase (black) when the material is sandwiched by the supporting substrate with a non-wet substrate with $\gamma_{l}=-0.4$ on the bottom.  The top surface is supported by the substrate with $\gamma_{u}=0$, which means a free boundary condition is imposed~\cite{Castro}.   The contact angle $\theta$ at the bottom substrate is determined from $\cos\theta=-0.4/0.042\simeq -10$, which gives the completely drying contact angle $\theta =\pi$.  Heterogeneous nucleation or wetting at the substrate is completely suppressed.

The contact angle at the neutral top substrate is given by $\theta=\pi/2$. We consider medium undercooling $\epsilon=0.2$ and low temperature $1/T\propto 1/\xi_{0}^{2}=60$. The periodic boundary condition is imposed along the horizontal $x$-axis. Similar to the homogeneous nucleation considered previously~\cite{Iwamatsu4}, the nucleation occurs inside the bulk.  The shape of the nucleus is almost circular and will grow isotropically with a seemingly constant interfacial velocity $v_{s}$. Therefore, the classic picture of homogenous nucleation and growth seems valid in this homogeneous-nucleation regime.

Figure \ref{fig:3} shows the time evolution of the volume fraction $f$ of transformed volume for two-dimensional $512\times512$ system with $\gamma_{l}=-0.4$ averaged over 50 samples. Since nucleation occurs uniformly as shown in Fig.~\ref{fig:2}, the sample-dependence of $f$ is small.  The standard deviations for $f$ at each time step is too small to be visible in Fig.~\ref{fig:3}.   The curve is very close to that of homogeneous nucleation in the infinite system, which is also averaged over 50 samples.  The curve for $\gamma_{l}=-0.4$, however, cannot be fitted to theoretical curve (\ref{eq:3-1}) by the least-square fitting of three parameters, $\tau_{\rm gr}$, $t_{\rm inc}$ and $m$ in principle because system is not homogeneous any more.  However it could be well represented by the KJMA formula (\ref{eq:3-1}) since the evolution of $f$ does not differ much from that of homogeneous nucleation. 

In fact, our simulation data in Fig.~\ref{fig:3} can be converted to the predicted KJMA linear relation (\ref{eq:3-3}) as shown in Fig.~\ref{fig:4}.  In order to convert the volume fraction $f$ to the Avrami plot, we have to specify the incubation time $t_{\rm inc}$ in (\ref{eq:3-3}).  Since we cannot fit the KJMA formula (\ref{eq:3-1}) directly to the simulation data for $f$, we cannot use least-square fitting to deduce $t_{\rm inc}$.  Instead, we have directly deduced incubation time from the averaged data of $f$ shown in Fig.~\ref{fig:3}.  We set $t_{\rm inc}=t_{0}$ where $t_{0}$ is the time step when the volume fraction $f$ start to increase from 0.  The transformed Avrami plot shows linear relation predicated from (\ref{eq:3-3}) which means that the classical KJMA formula (\ref{eq:3-1}) seems to be correct in this homogeneous nucleation regime.  The exponents $m$ deduced from this Avrami plot are almost the same for different $\gamma$ and are 
$m=$4.36, 4.26 and 3.86 for $\gamma=0.0$, -0.02 and -0.4 respectively, which are very close to the $m=$4.45 for the homogeneous nucleation.  These values, however, are larger than the ideal value $m=3$ calculated from (\ref{eq:3-9a}) when $d=2$.

\subsection{Heterogeneous nucleation regime}

In contrast to the homogeneous nucleation regime in Fig.~\ref{fig:2}, a new nuclear embryo preferentially appears near the bottom substrates and the new phase starts to grow from the bottom substrates in the heterogeneous nucleation regime.  When the surface field $\gamma_{l}$ is weak, only a small number of nuclei is formed at the lower substrate and grow as shown in Fig.~\ref{fig:5}. Initially we observe semicircular nucleus nucleated on the bottom substrate.  Eventually they grow with constant velocity $v_{s}$ given by (\ref{eq:3-9}) and start to coalesce.   Then the evolution of the volume fraction $f$ shows large sample-dependence because it depends strongly on when and how much the nuclei are formed.  

\begin{figure}[htbp]
\begin{center}
\includegraphics[width=0.85\linewidth]{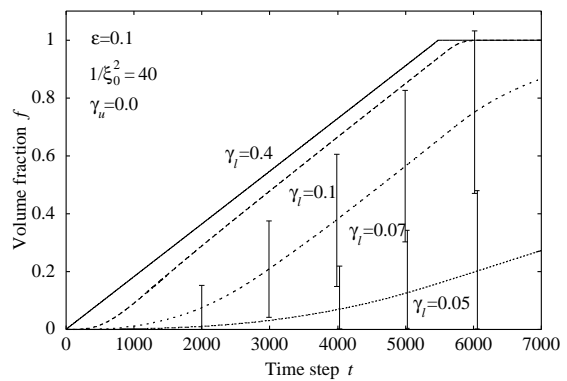}
\end{center}
\caption{
The same as Fig.~\ref{fig:3} but with a weakly wet substrate with $\gamma_{l}$=+0.4, 0.1, 0.07 and 0.05 at the bottom for the low undercooling $\eta$=0.1 and at a medium temperature $1/\xi_{0}^{2}=40$. The standard deviations are indicated only for $\gamma_{l}=0.07$ and $0.05$ bu vetical lines.  They are too small to be shown for $\gamma_{i}=0.1$ and $0.4$.}
\label{fig:6}
\end{figure}

\begin{figure*}[htbp]
\begin{center}
\includegraphics[width=0.8\linewidth]{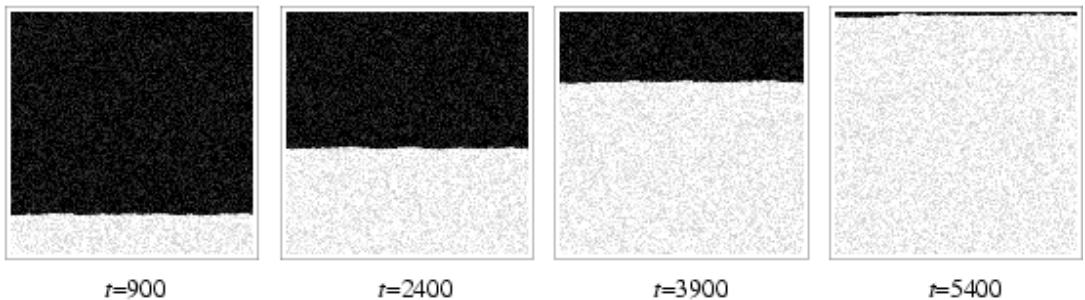}
\end{center}
\caption{
The same as Fig.\ref{fig:2} but with a strongly wet substrate with $\gamma_{l}=+0.4$ at the bottom for the low undercooling $\eta=0.1$ and at a medium temperature $1/\xi_{0}^{2}=40$.  Homogeneous nucleation is suppressed and nucleation occurs preferentially from the lower wet substrate. }
\label{fig:7}
\end{figure*}

Since there is large scattering of data for the evolution of volume fraction $f$, we show in Fig.~\ref{fig:6} the time evolution of $f$ averaged over 50 samples. There are large standard deviations for $\gamma_{l}=0.05$ and $0.07$ which could be as large as the average $f$ itself.  This is because there is large scattering in the time when the nucleation starts and in the number of nuclei formed.  It should be noted that the complete wetting condition is satisfied if $\gamma_{l}>0.042$ as the surface tension of the nucleus is given by $\sigma\simeq 0.042$.  As we increase the strength $\gamma_{l}$ of the surface field, the scattering of data decreases as the nucleation occurs almost uniformly on the substrate as shown in Fig.~\ref{fig:7}.  The standard deviation for $\gamma_{l}=0.1$ and $0.4$ becomes less than 10\% of $f$ itself and have not been shown in Fig.~\ref{fig:7}.

When the surface field is strong, heterogeneous nucleation occurs easily, and there are numerous nuclei formed at the substrate.  Initially we observe a semicircular nucleus nucleated on the bottom substrate.  They soon coalesce and will grow with constant velocity.  Finally the linear horizontal moving front of transformed phase appears and moves upward with constant velocity $v_{s}$ (Fig.~\ref{fig:7}).  

Plane-wave like moving front reflects in the non-KJMA behavior of the volume fraction $f$ of transformed volume for $\gamma_{l}=0.4$ shown in Fig.~\ref{fig:6}.  When the substrate is strongly attractive (wet) with $\gamma=0.4$, the volume fraction $f$ increase linearly as the function of time step.  In this case, the contact angle $\theta$ at the bottom substrate is determined from $\cos\theta=0.4/0.042\simeq 10$, which gives complete wetting contact angle $\theta =0$.  The front velocity estimated from (\ref{eq:3-9}) gives $v_{s}=0.075$ as $D=1/2$ and $\epsilon=0.1$, which nicely corresponds to the value $v_{s}=512/5500\simeq 0.093$ estimated from the slope $1/5500$ of the straight line for $\gamma_{l}=0.4$ in Fig.~\ref{fig:6}.

Therefore, in this regime, an Avrami plot (\ref{eq:3-3}) does not give any information about the microscopic picture of growth kinetics.  It is also clear from Fig.~\ref{fig:6} that when the substrate is very attractive ($\gamma_{l}=0.4$) the phase transformation is accomplished within a relatively short time compared to the weakly attractive case ($\gamma_{l}=0.05$).  In the latter case, the bottom substrate acts to enhance the nucleation in the bulk, but the phase transformation still looks like homogeneous nucleation in Fig.~\ref{fig:6} within the bulk. 

\subsection{Homogenous-Heterogeneous coexistence regime}

\begin{figure*}[htbp]
\begin{center}
\includegraphics[width=0.8\linewidth]{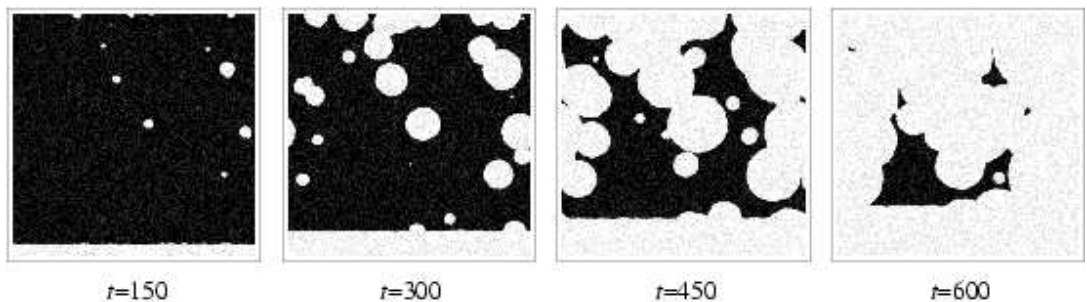}
\end{center}
\caption{
The same as Fig.~\ref{fig:2} but with a strongly wet substrate with $\gamma_{f}=0.4$ at the bottom for a medium undercooling $\epsilon=0.2$ and at a lower temperature $1/\xi_{0}^{2}=80$.  Initially the nucleation starts from the bottom substrate as the heterogeneous nucleation proceeds.  Later, the homogenous nucleation and growth within the bulk starts.  Eventually, both the homogeneous and the heterogeneous nucleation and growth coexist.}
\label{fig:8}
\end{figure*}

When the undercooling is moderately high and the temperature is low, homogeneous nucleation and heterogeneous nucleation can coexist if the surface field is attractive and strong.  Figure \ref{fig:8} show a typical evolution pattern when the surface field is $\gamma_{l}=0.4$.  Since the temperature is low, the incubation time for homogeneous nucleation is long.  Also, since the undercooling is medium, the growth velocity is slow.  Then the heterogeneous nucleation and subsequent growth is not shadowed by the homogeneous nucleation.

\begin{figure}[htbp]
\begin{center}
\includegraphics[width=0.85\linewidth]{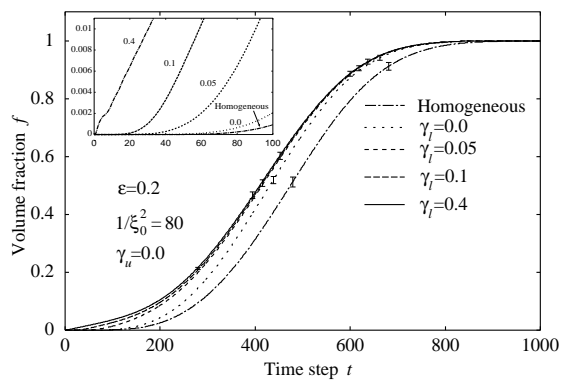}
\end{center}
\caption{
The same as Fig.~\ref{fig:3} but with a wet substrate with $\gamma_{l}$=+0.4, 0.1, 0.05 and 0.00 at the bottom for the medium undercooling $\eta$=0.2 and at a low temperature $1/\xi_{0}^{2}=80$.  When the surface field $\gamma_{l}$ increases, heterogeneous nucleation with a low energy barrier appears.  Then nucleation starts earlier.  The global shape of the transformation curve $f$ does not differ much from that of KJMA prediction for the homogeneous nucleation.  However, the incubation time $t_{\rm inc}$ becomes apparently shorter. Vertical lines indicate the standard deviations.
}
\label{fig:9}
\end{figure}

The corresponding evolution curve of volume fraction $f$ looks superficially similar to the KJMA prediction for the homogeneous nucleation and growth.  However, the incubation times become apparently shorter as we increase the surface field $\gamma_{l}$ (Fig.~\ref{fig:9}), which means that the phase transformation starts earlier because the incubation time for the heterogeneous nucleation becomes shorter from (\ref{eq:3-4}) and (\ref{eq:3-10}).

\begin{figure}[htbp]
\begin{center}
\includegraphics[width=0.85\linewidth]{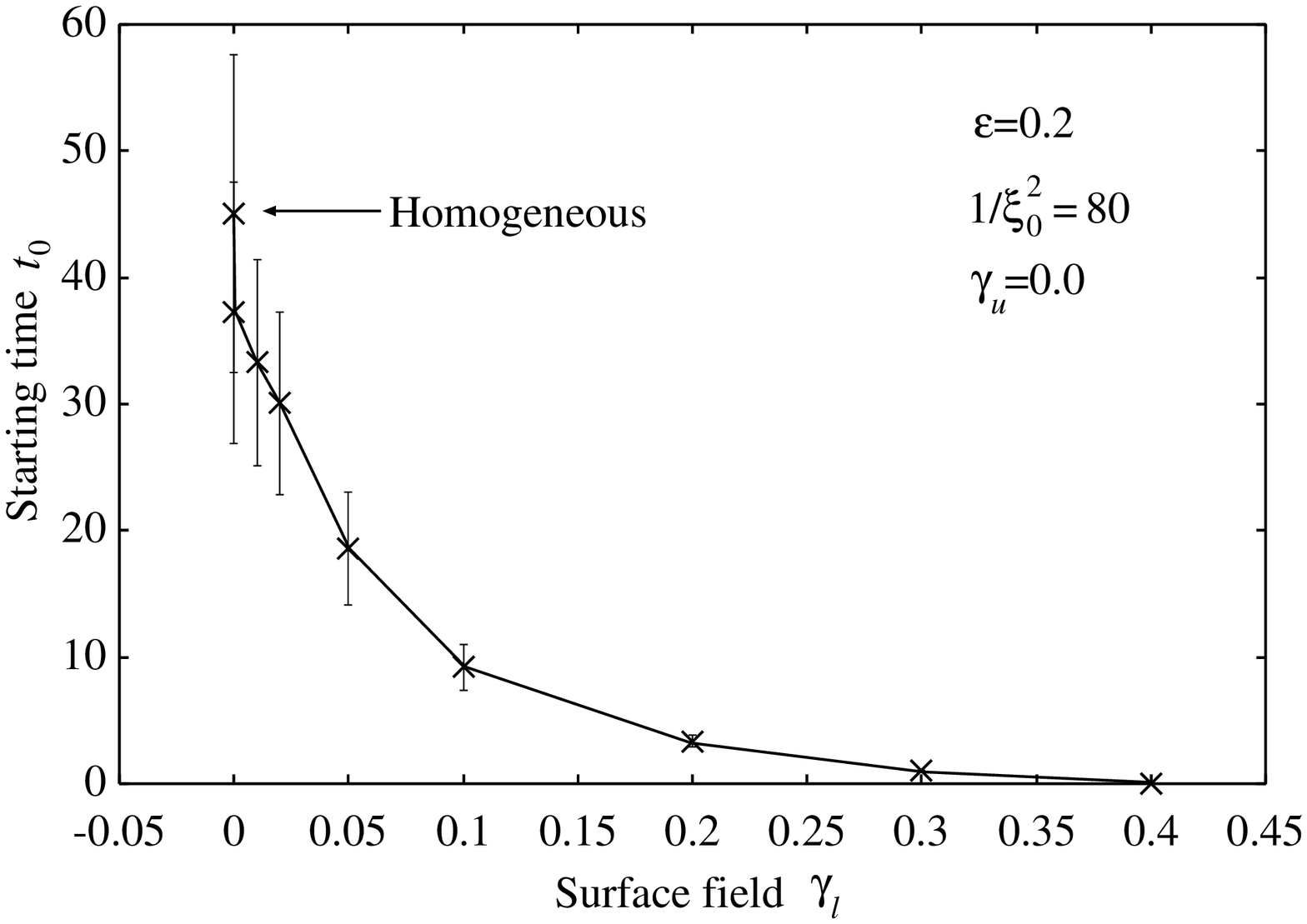}
\end{center}
\caption{
The start time $t_{0}\simeq t_{\rm inc}$ of the phase transformation averaged over 50 samples estimated from the evolution of the volume fraction $f$ as the function of the surface field $\gamma_{l}$ when $\epsilon=0.2$ and $1/\xi_{0}^{2}=80$. Vertical lines indicate the standard deviation.  }
\label{fig:10}
\end{figure}

Figure \ref{fig:10} shows the starting time $t_{0}$ of phase transformation as the function of the surface field $\gamma_{l}$.  Since we cannot use the KJMA formula (\ref{eq:3-1}) anymore to fit the simulation data to deduce the incubation time $t_{\rm inc}$, we have used starting time $t_{0}$ instead.  The figure clearly indicates the expected decrease of starting time $t_{0}$ as we increase the surface field $\gamma_{i}$.  Therefore, the phase transformation can start earlier due to the presence of the wet substrate.

\begin{figure}[htbp]
\begin{center}
\includegraphics[width=0.85\linewidth]{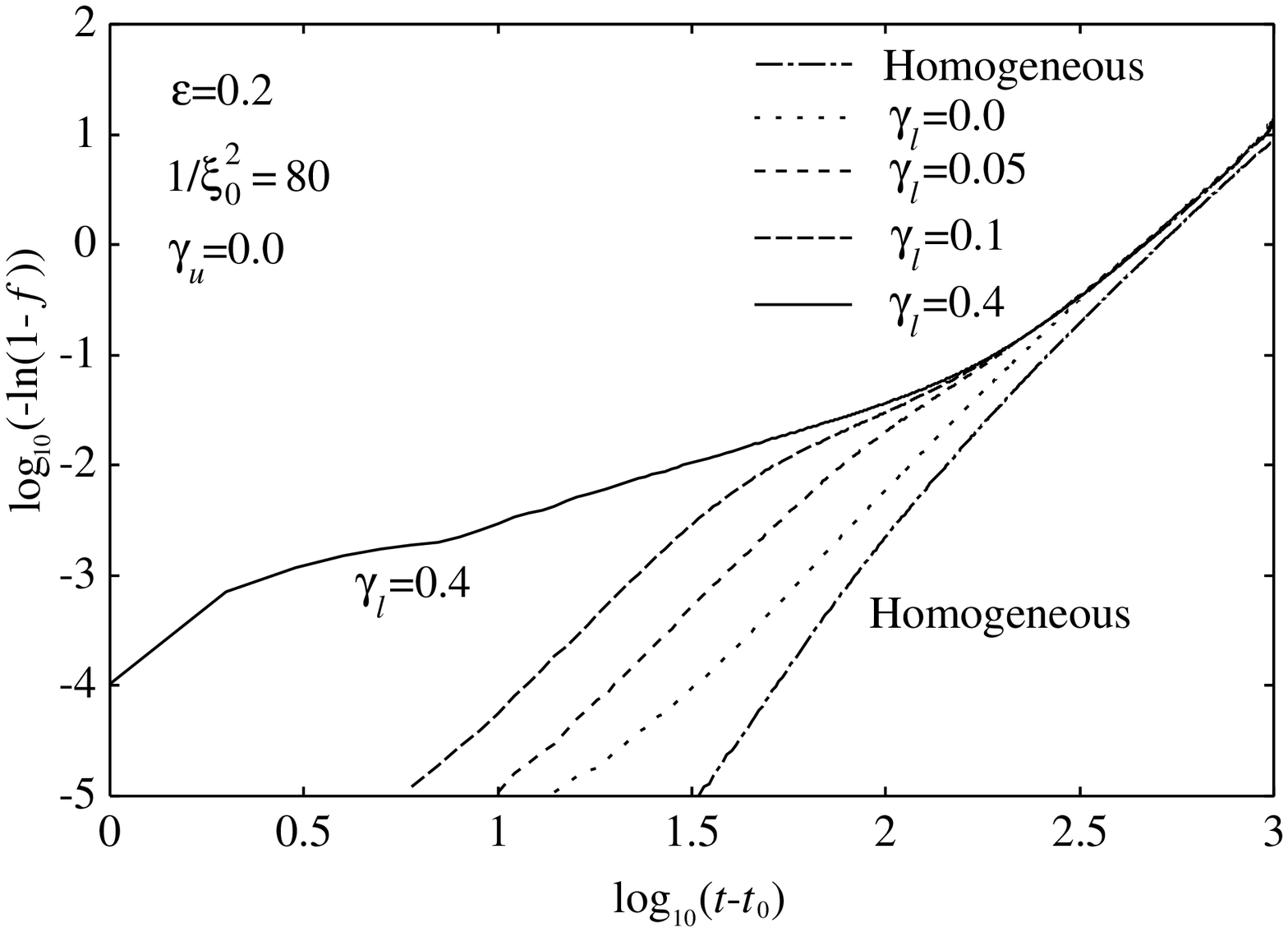}
\end{center}
\caption{
The Avrami plot corresponding to Fig. \ref{fig:9}. Although the curves for the volume fraction $f$ all look very similar to the curve for homogeneous nucleation in Fig.\ref{fig:9}, the Avrami plots clearly show the transition from one-dimensional growth at earlier time with a gentle slope with smaller exponent $m\simeq 1.06$ to the two-dimensional growth at later time with a steep slope with larger exponent $m\simeq 2.94$ when the surface field is strongly attractive ($\gamma_{l}=0.4$). 
}
\label{fig:11}
\end{figure}

The Avrami plots which correspond to Fig.~\ref{fig:9} are shown in Fig.~\ref{fig:11} where we have used the starting time $t_{0}$ estimated from Fig.~\ref{fig:7} instead of the incubation time $t_{\rm inc}$ in (\ref{eq:3-3}).  Although the curves in Fig.~\ref{fig:9} looks similar to the KJMA prediction, the Avrami plot in Fig.~\ref{fig:11} clearly display a large deviation from the KJMA prediction of linear relation in early stage of growth.  When this surface field is strongly attractive ($\gamma_{l}=0.4$), in particular, the Avrami plot consists of two linear segments with different Avrami exponents $m$.  In the early stage of transformation $m\simeq 1.06$ while in later stage it becomes $m\simeq 2.94$, which is very close to the ideal value $m=3$ for two-dimensional isotropic growth predicted from (\ref{eq:3-9a}).

Therefore, nucleation at the early stage is dominated by heterogeneous continuous nucleation in 0 dimensions or the site saturation in 1 dimension with Avrami exponent $m=1$ from (\ref{eq:3-9b}).  This contribution from heterogeneous nucleation to the transformed volume $f$ is soon suppressed due to the exhaustion of the nucleation site on the bottom substrate.  At a later stage, heterogeneous nucleation is shadowed by homogeneous continuous nucleation which starts inside the bulk material since there is still much of the site to be use for homogeneous nucleation to start.  Then the Avrami exponent in later stage becomes close to the ideal value $m=3$ predicted from (\ref{eq:3-9a}) for the $d=2$ dimensional continuous nucleation.

\section{Conclusion}
\label{sec:sec4}

In this study, we used the cell dynamics method~\cite{Iwamatsu1} to study the possible scenario of heterogeneous nucleation and growth of new phase in unified manner within the framework of the time-dependent Ginzburg-Landau (TDGL) model.  We found that the formula for the volume fraction of transformed volume due to Kolmogorov-Johnson-Mehl-Avrami (KJMA) theory and the so-called Avrami plot do not necessarily capture the essential future of the dynamics of heterogeneous nucleation.  In particular, when the heterogeneous nucleation is dominated, the Avrami plot is totally useless.  Rather, a simple linear plot of the transformed volume $f$ against time could give a linear relation between the transformed volume and time, and the front velocity directly gives the interfacial velocity of phase transformation. 
Therefore, a naive interpretation of Avrami plot when the supporting substrates are changed, needs careful examination~\cite{Jeong,Oshima}.

We have also predicted that, in general, heterogeneous nucleation and homogeneous nucleation could coexist.  Then the simple Avrami plot would consist of two linear segments with inflection.  The Avrami exponents $m$ of two segments would give the dominant mechanism of phase transformation.  It is predicted that since the heterogeneous nucleation could occur earlier, we might have a smaller Avrami exponent an at early stage and a larger Avrami exponent at a later stage.

Since the TDGL model with the cell dynamics method is flexible and computationally simple, yet it has a direct connection to the thermodynamics of the system considered, it can be used to study various scenarios of the dynamics of phase transformations. In fact it has already applied to those materials with a complex phase diagram~\cite{Iwamatsu3}.  Although standard cellular automaton~\cite{Marx,Hesselbarth} and the lattice model ~\cite{Rollett} could also be used to simulate the complex dynamics of phase transformation, they have to introduce an artificial evolution processes algorithmically. On the other hand, the evolution in our cell dynamics method is driven by the free energy and the thermal noise, and it is completely free from artificial parameters.  Therefore, our TDGL model with cell dynamics method is more natural and easy to use in the study of various phase transformation scenarios, in particular, of the time evolution including the incubation time without introducing many unknown and uncontrollable parameters.  


\begin{thebibliography}{99}
\bibitem{IwamatsuE} A factor 1/2 is misssing in Eq.(12) of reference [13].
\end{thebibliography}
\end{document}